\magnification=\magstep1
\baselineskip=14pt
\input epsf.tex

\centerline{\bf ALGORITHMIC RANDOMNESS, PHYSICAL ENTROPY, MEASUREMENTS,}
\centerline{\bf AND} 
\centerline{\bf THE DEMON OF CHOICE}
\bigskip
\centerline {W. H. Zurek}
\centerline {Theoretical Division}
\centerline {T-6, MS B288}
\centerline {Los Alamos National Laboratory}
\centerline {Los Alamos, NM 87545}

\bigskip
\centerline {ABSTRACT}
\bigskip
Measurements --- interactions which establish correlations between a system and 
a recording device --- can be made thermodynamically reversible. One might 
be concerned that such reversibility will make the second law of thermodynamics 
vulnerable to the designs of the {\it demon of choice}, a selective version of 
Maxwell's demon. The strategy of the demon of choice is to take advantage 
of rare fluctuations to extract useful work, and to reversibly undo 
measurements which do not lead to such a favorable but unlikely outcomes. 
I show that this threat does not arise as the demon of choice cannot operate 
without recording (explicitely or implicitely) whether its measurement was 
a success (or a failure). Thermodynamic cost associated with such a record 
cannot be, on the average, made smaller than the gain of useful work derived 
from the fluctuations.

\vfill
\centerline{ \copyright {by Wojciech Hubert Zurek, March 12 1998}}
\eject

When I was asked to write for a volume dedicated to Richard Feynman, I 
decided that I should select the subject in which I was influenced by him the 
most, and which would still be consistent with the overall theme of computation 
and physics. And these influences started well before I met him in 
person: I got Feynman's ``Lectures on Physics'' more than a quarter century ago,
in Polish translation, from my father. As a finishing high school student 
I was accompanying him on a hunting expedition in the lake district of Poland
--- a remote corner of the country. Every few days we drove 
for supplies to the provincial capital, and there I noticed the volumes in 
the local bookstore. My father asked why (the expense was considerable), but 
surprisingly easily gave way to my arguments. I spent much of the rest of 
the hunting vacation (a couple of weeks altogether) getting through volume I. 

Over the years I have developed a habit of treating the ``Lectures'' sort of 
like a collection of poems. I like some ``poems'' more than others, and I return
to the favorites now and again. And when I am stuck with a physics problem, 
reading a few of the relevant ``poems'' is often the best way to get 
``unstuck''. But there are a few chapters which have been read over and over 
again without any such an ulterior motive, for sheer pleasure. Amongst them, 
I would certainly include the discussion of the fluctuations and the second law 
(the famous ``ratchet and pawl'' argument$^1$).

Thermodynamic concerns and arguments have often pre-saged the deepest 
developments in physics. I suspect this is because thermodynamics ``knows'' 
about the physical relevance of information, and hence, it knew about the Planck
constant, stimulated emission, black hole entropy, and so on. When I met 
Feynman in person for the first time (at a small workshop organised near Austin,
Texas, by John Archibald Wheeler in the Spring of 1981), I remember --- amongst 
other things ---  a thermodynamic argument he used to great effect to prove 
that one cannot accelerate elementary particles by shaking them together with a 
bunch of heavier objects, so that they could acquire equipartition kinetic 
energies (and therefore, because of their small mass, enormous momenta).  This 
idea (credible at first sight, as it is akin to the Fermi acceleration of 
cosmic rays) was brought up by one of the participants. It would not work --- 
Feynman argued  --- because all sorts of other modes of the vacuum would have 
to get their fair share of energy, creating an equilibrium heat bath, with 
approximate equipartition between all the modes (rather then with the energy 
in the elementary particles one really wanted to accelerate in 
the first place). 

But that was not the most vivid memory of that first encounter with the man 
whose ``Lectures'' I had acquired a decade or so earlier. Rather, I remember 
best that he showed up at the first lecture unshaved and uncombed, with dry 
grass in his hair. It turned out that he spent the night outside --- apparently,
he decided the accomodations for the speakers (which were in the posh tennis 
club) were too opulent, returned the key to his apartment at the reception, and
decided to ``camp out''. During the morning coffee he has also reported in 
detail (and with great gusto) how he had trouble breaking the code to get into 
his briefcase (where he had the sweater --- it got cold). He knew the code, of 
course, by heart, but it was middle of the night, so he somehow had to dial it 
in complete darkness. He clearly relished the challenge. I do not remember how
did he solve the problem, but the flavor of the adventure and of his report
was very much in the spirit of the ``adventures of a curious character''.
And all of this was a few months after his (first) cancer operation. 

I came to talk to Feynman regularly, more or less once a month, during my Tolman
Fellowship at Caltech (which started in the Fall of 1981), and a bit less often 
for a few years afterwards. I have also sat occasionally in the class on physics
and computation he taught with John Hopfield. And I remember discussing with 
him (among other subjects) the connection between physics, information and 
computation. In fact, this was a recurring theme. For me, it became somewhat 
of an obsession early on --- I really liked the universality of Turing machines,
the halting problem, and the algorithmic view of information. While I was in 
Austin the fascination with these ideas and their possible relevance for physics
was reinforced under the influence of John Wheeler. Which brings me, at long 
last, to the {\it algorithmic information content}, {\it measurements}, 
and various {\it thermodynamic demons} which probe the utility of acquired 
information.

Maxwell's demon --- a hypothetical intelligent entity capable of performing 
measurements on a thermodynamic system and using their outcomes to extract 
useful work --- was considered a threat to the validity of the second law of 
thermodynamics for over a century.$^{2,3}$ Feynman was fascinated with the 
subject, and his discussion of ratchet and pawl$^1$ banished forever the 
``unintelligent'' trapdoor version of the demon by clarifying and updating 
the influential argument put forward by Smoluchowski$^4$ much earlier, and in 
a rather different setting.\footnote*{Smoluchowski's original trapdoor was 
a hole surrounded by hairs combed so that they all come out on the same side of 
the partition between the two chambers (rather than a real trap{\it door}). 
Naively, this arrangement of hairs should favor molecules passing in the
direction in which the hair is combed, and impede the reverse motion. 
Smoluchowski pointed out that thermal fluctuations will ``ruffle the hair''
and make this arrangement ineffective as a rectifier of fluctuations when the whole system is 
at the same fixed temperature. Numerical simulations of trapdoors confirm these 
conclusions$^5$. They also show why our intuition based on far-from-equilibrium 
behavior of trapdoors can be easily misled.} However, Smoluchowski's trapdoor 
carries out no (explicit) measurements. Therefore, trapdoors and ratchets and 
pawls can be analysed without reference to information.$^{1,4}$

The complete Maxwell's demon should be able to measure, and it (...?; he? she?!)
should be of course intelligent. Smoluchowski's trapdoor does not fit this bill.
Measurements were incorporated into the discussion by Szilard$^6$, Landauer$^7$,
and Bennett$^{8}$ who have argued, in a setting involving ensembles of demons, 
that the acquisition of information is only possible when the demon's memory 
is repeatedly erased, to prepare it for the new data. The {\it cost of erasure}
eventually offsets whatever thermodynamic advantages of the demon's information 
gain might offer. This point (which has come to be known as 
``Landauer's principle'') is now widely recognised as a key ingredient of 
thermodynamic demonology. This originally classical reasoning has been since 
extended to quantum physics$^{9,10}$, and may even be 
experimentally testable.$^{11}$

However, the widespread fascination with Maxwell's demon is ultimately due to 
its intelligence. A demon will record a specific outcome of the measurement, and
--- using its intelligence --- will try to make an optimal decision about the
best possible action, which would maximize the work extracted from a given
recorded phase space configuration. This is very much the course of action we
take (although, fortunately for us, in a far-from-equilibrium setting). How can 
one convince an intelligent demon that, all cleverness notwithstanding, its 
attempts at defeating the second law are doomed? This is hard to accomplish at 
the level of ensembles: Each demon knows nothing but its own record, and need 
not care about the other members of ``its ensemble'' that have found out 
something else in their measurements --- it will find out solution to its own 
problem.

The ultimate analysis of Maxwell's demon must involve a definition of 
intelligence, a characteristic which has been all too consistently banished 
from discussions of demons carried out by physicists. On the other hand, 
intelligence has been --- since Turing and his famous test --- often invoked 
in the discussions of computer scientists. To convince ourselves (and the 
intelligent demon) of the limits imposed by the second law we shall, following
Ref. 12, adopt an operational definition of intelligence which arose in the 
context of the theory of computation. It is based on the so-called Church--Turing 
thesis$^{13}$ --- which in effect formalizes Turing's expectations about the 
``mental'' capabilities of computers and states that intelligence is equivalent 
to the same kind of information processing that is in principle implementable on
a universal computer.

Using the Church--Turing thesis as a point of departure, the present author 
has demonstrated that even this intelligent threat to the second law can be 
eliminated --- the original ``smart'' Maxwell's demon can be exorcized. This
is easiest to establish when one recognizes that the net ability of demons 
to extract useful work from systems depends on the sum of measures of 
two distinct aspects of disorder:$^{12}$ 

(i) The usual {\it statistical entropy} given by:
$$H(\rho) = - Tr \rho \lg \rho \eqno (1)$$
where $\rho$ is the density matrix of the system, determines the ignorance
of the observer.

(ii) The {\it algorithmic information content:}$^{14-20}$
$$K(\rho) = | p^*_{ \rho } | \eqno (2)$$
is given by the size (``$|...|$''), in bits, of the shortest algorithm ($p^*$)
which, for an ``operating system" of a given Maxwell's demon, can reproduce 
the detailed description $(\rho)$ of the state of the system. $K(\rho)$ 
quantifies the cost of storing of the acquired information, which is related to 
the randomness inherent in the state of the system revealed by the measurement.

The Church--Turing thesis enters in this second algorithmic ingredient, 
as it involves an assumption that the intellectual abilities of 
Maxwell's demons can be regarded as equivalent to those of a universal 
Turing machine: It is assumed that demons can execute programs 
(such as $p^*_{\rho}$) to reconstruct records of past 
measurements out of their optimally compressed versions, or to carry out other 
logical operations in optimizing performance. Algorithmic information content 
provides a well-defined measure of the storage space required to register 
the known characteristics of the system.

{\it Physical entropy}$^{12}$ is the sum of the statistical entropy and of
the algorithmic information content:
$${\cal Z} (\rho) = H(\rho) + K(\rho) \eqno (3)$$
Above, it is assumed that the base for the logarithm in Eq. (1) is the
same as the size of the alphabet used by the computer which constitutes the
operating system of the Maxwell's demon. In practice, it is customary and
convenient to employ a binary alphabet, so that both $H(\rho)$ and
$K(\rho)$ are measured in bits.

In order to appreciate the physical significance of the algorithmic randomness 
contribution, it is useful to discuss the behavior of $H,\ K$ and ${\cal Z}$ in 
the course of measurements and to follow the operations of the engines 
controlled by demons. In short, the two measures turn out to be complementary
--- not in the quantum sense, but a bit like kinetic and potential energy ---
and their sum is, on the average, conserved under optimal measurements carried 
out on an equilibrium ensemble. Analysis which leads to this conclusion was 
carried out by this author$^{12,10}$ and extended by Caves$^{21}$. Below we 
offer only a brief summary of the salient points.

In the course of ideal measurement on an equilibrium ensemble the decrease of 
ignorance is, on the average, compensated for by the increase of the size of 
the minimal record:$^{12}$
$$\Delta H \simeq -  < \Delta K >. \eqno (4)$$
Consequently, physical entropy ${\cal Z}$ plays a role analogous to a constant 
of motion. The transformation of the state of the system is now, however, 
brought about by a {\it demonical} (rather than {\it dynamical}) evolution, 
by the act of acquisition of information. This ``conservation law" can be 
demonstrated within the context of the algorithmic theory of 
information$.^{12,10,21,22}$ However, its validity can be traced to coding 
theory:$^{12,21-23}$ According to the noiseless coding theorem of 
Shannon,$^{23}$ the minimal size ${\cal L}$ of the message required to encode 
information which corresponds to a decrease
of entropy by $\Delta H$ is, on the average over all of the messages,
bounded by:
$$\Delta H \leq {\cal L} < \Delta H + 1 $$
This inequality is used in the proof of Eq. (4) and is ultimately
responsible for the constancy of the physical entropy $\cal Z$ in the
course of the measurement.$^{12,21}$

The role of $\cal Z$ in determining the efficiency of demon-operated
engines is the ultimate reason for regarding $\cal Z$ as physical entropy.
For, the total amount of work which can be extracted from a physical system
in contact with a heat reservoir
of temperature $T$ in the course of a cycle which involves a measurement
$(\rho \rightarrow \rho_i)$ and isothermal expansion $(\rho_i \rightarrow
\rho)$ can be made as large as, but no larger than:$$\Delta W = k_BT
({\cal Z}(\rho) - {\cal Z}(\rho_i)) \eqno (5)$$

To justify this last assertion, I shall appeal to Landauer's principle$^7$ 
which formalizes earlier remarks of Szilard$^6$ and states that erasure of one 
bit of information from the memory carries a thermodynamic price of $k_BT$. 
Although Landauer's principle assigns a definite price to the storage of
information, this price need not be paid right away: a demon with a large
unused memory can continue to carry out measurements as long as it has room
to store information. However, such a demon poses no threat to the
second law: its operation is not truly cyclic. In effect, it operates by
employing its initially empty memory as a low temperature (zero entropy)
heat sink.

Erasure of the results of used up measurements carries a price tag of
$$\Delta W^- = T<(K(\rho_i) - K(\rho)> \ , 
\eqno (6a)$$
which must be subtracted from the gain of useful work
$$\Delta W^+ = T(H(\rho) - H(\rho_i)) \ , \eqno (6b)$$
to obtain the net work extracted by the demon. This immediately justifies 
Eq. (5).  The hybrid $\cal Z$ is the physical entropy which provides the
demon with an individual, personal measure of the potential for thermodynamic 
gains due to the information in its possession. It also demonstrates that 
a demon operating on a system in thermodynamic equilibrium will never be able 
to threaten the second law, for the ensemble average of $\cal Z$ is at best 
conserved, so that $< \Delta {\cal Z} > \leq 0$
in course of the process of acquisition of information.

This last assertion is, however, justified only if the demon is forced to
complete each measurement-initiated cycle. One can, by contrast, imagine a
{\it demon of choice}, an intelligent and selective version of Maxwell's demon, 
who carries out to completion only those cycles for which the initial state of 
the system is sufficiently nonrandom (concisely describable, or 
{\it algorithmically simple}) to allow for a brief compressed record
(small $K(\rho))$. This strategy appears to allow the demon to extract 
a sizeable work $(\Delta W^+)$ at a small expense 
$(\Delta W^-)$. Moreover, if the measurements can be reversibly undone, 
then the ones with disappointing outcomes could be reversed at no cost. 
Such demons would still thereaten the second law, even if the threat is
somewhat more subtle than in the case of Smoluchowski's trapdoor.

Caves$^{22}$ has considered and partially exorcised such a demon of choice 
by demonstrating that in any case the net gain of work cannot exceed $k_BT$ 
per measurement. Thus, the demons would be, at best, limited to exploiting 
thermal fluctuations. Moreover, in a comment$^{24}$ on Ref. 22 it was noted that
taking advantage of such fluctuations is not really possible. Here I shall 
demonstrate that the only decision making process free of inconsistencies
necessarily leaves in the  observer's (demon's) memory a ``residue'' which 
requires eventual erasure. The least cost of erasure of this residue is just 
enough to restore the validity of the second law. The aim of this paper is to 
make this argument (first put forward by this author at the meeting of the 
{\it Complexity, Entropy, and the Physics of Computation} network of the 
Santa Fe Institute in April of 1990) more carefully and more precisely.

To focus on a specific example consider a {\it Gabor's engine}$^{25}$
illustrated in Fig. 1. There, the unlikely but profitable
fluctuation occurs whenever the gas molecule is found in the small
compartment of the engine. The amount of extractable work is:
$$\Delta W^+_p = k_BT ~ \lg(L/\ell) \eqno (7)$$
The expense (measured by the used up memory) is only:
$$\Delta W^- = k_BT  , \eqno (8)$$
so that the net gain of work per each successful cycle is:
$$\Delta W _p = k_BT ~ ( \lg( L/\ell) - 1) \eqno (9)$$
The more likely ``uneconomical" cycles would allow a gain of work:
$$\Delta W^+_u = k_BT ~ \lg L/(L-\ell) \ ,  \eqno (10)$$
so that the cost of memory erasure (still given by Eq. (8))
outweighs the profit, leaving the net gain of work:
$$\Delta W_u = - k_BT(1 - \lg L/(L-\ell)). \eqno (11)$$
When each measurement is followed by the extraction and erasure routine,
the averaged net work gain per cycle is negative (i.e., it becomes a
loss):
$$< \Delta W > = {\ell \over L} \Delta W_p + {L-\ell \over L} \Delta W_u
= - k_BT [ 1 + ({\ell \over L} \lg {\ell \over L} + {L - \ell \over L} \lg
{L - \ell \over L})] \eqno (12)$$
The break even point occurs for the case of Szilard's engine$^{6}$, where
the partition divides the container in half.
In the opposite limit, $\ell /L << 1$, almost every measurement leads to an
unsuccessful case which results in a negligible amount of extracted work
but undiminished cost of erasure per cycle.

The design of the demon of choice attempts to capitalize on precisely this 
otherwise unprofitable limit by {\it undoing} all of the likely (and
unprofitable) measurements at no thermodynamic cost, thus avoiding the necessity
for erasure of the unused outcomes. It is important to emphasize that a
measurement of the thermodynamic quantities can be indeed undone at no cost: 
A prejudice that measurement must be thermodynamically expensive goes back at 
least to the ambiguities in the original paper of Szilard$^{6}$ (who
has hinted at, but failed to clearly identify erasure as the only
thermodynamically expensive part of the measuring process), and was further
reinforced by the popular (but incorrect) discussion of Brillouin.$^{26}$
Figure 2 demonstrates how to carry out a measurement on a particle in the 
Gabor's engine (such measurement becomes reversible when the operations 
indicated are carried out infinitezimally slowly).

The purpose of the measurement is to establish a correlation between the state 
of the system and the record --- the state of the few relevant bits of memory. 
In the context of this paper we shall focus on the measurements which correlate 
memory with a cell in the phase space or a subspace of the Hilbert space of the system (corresponding to the projection operator $P_i$). In concert with the 
usual requirements I shall demand that the collection $\{ P_i \}$ of all the
measurements be mutually exclusive $(Tr(P_i, P_j) = 0)$, and exhaustive 
$(\Sigma_i P_i = 1)$. To avoid problems associated with quantum measurements we 
shall also demand that the measured observables should commute with the density 
matrix of the measured system $[P_i, \rho_S] = 0$. Thus, we shall allow for the 
best case$^9$ (from the demon's point of view), with no additional thermodynamic
inefficiencies associated with the reduction of the state vector introduced 
into quantum measurement through decoherence.$^{28-31,10,11}$

A measurement performed by the demon, when viewed from the outside, results
in the correlation between the state of the system (i.e. location of 
the particle in the Gabor's engine) and the state of the demon's memory. 
The total entropy can be prevented from increasing, as the only requirement 
for a successful measurement is to convert initial density
matrix of the combined system-demon:
$$\rho_{SD}^{(o)} = \rho_S \times \rho_D^{(o)} = (\Sigma_i p_i P_i) \times
\rho_D^{(o)} \eqno (13a)$$
into the correlated$^{9,10,28-31}$:
$$\rho_{SD} = \Sigma_i p_i(P_i \times \rho_D^{(i)}) \eqno (13b)$$
Above, we have implicitly assumed that the measurement is exhaustive in the 
sense that the further refinements will reveal uniform probability 
distribution within the partitions defined by $P_i$. This need not be the case 
--- it is straightforward to generalize the above formulae to the case when
the different memory states of the demon are correlated with density matrices
of the system. In any case, the entropies of $\rho_D^{(i)}$ and $\rho_D^{(o)}$ 
can, in principle, be the same: For, there exists a unitary 
{\it controlled-not} - like evolution operator:
$$U = \Sigma_i P_i \times (\vert \delta_i >< \delta_o \vert + \vert
\delta_o >< \delta_i \vert) \eqno (14)$$
with $\vert \delta_i>$ and $\vert \delta_o>$ defined by $\rho_D^{(i)} =
\vert \delta_i >< \delta_o \vert \rho_D^{(o)}$, providing that
$\rho_D^{(i)}$ correspond to distinguishable (orthogonal) memory states
of the demon --- a natural requirement for a successful measurement.

The statistical entropy of the system-demon combination is obviously the
same before and after measurement, as, by construction of $U$, 
$H(\rho_D^{(i)}) = H(\rho_D^{(o)})$. Moreover, the measurement is obviously 
reversible: Applying the unitary evolution operator, Eq. (14), twice,
will restore the pre-measurement situation.

From the viewpoint of the outside observer, the measurement leads to a
correlation between the system and the memory of the demon: The ensemble
averaged increase of the ignorance about the content of demon's memory;
$$\Delta H_D = H(\rho_D) - H(\rho_D^{(o)}) = - \Sigma_i  p_i \lg p_i \ , 
\eqno (15)$$
(where $\rho_D= Tr_S \rho_{SD}$ and $H(\rho) = - Tr \rho \lg \rho$) is
compensated for by the increase of the mutual information defined as;
$$I_{SD} = H(\rho_D) + H(\rho_S) - H(\rho_{SD}), \eqno (16)$$
so that $\Delta H_D = \Delta I_{SD}$ (see Refs. 29 and 33 for the Shannon 
and algorithmic versions of this discussion in somewhat different settings).

From the viewpoint of the demon the acquired data are definite: The outcome is some
definite demon state $\rho_D^{(n)}$ corresponding to the memory state $n$, and
associated with the most concise record --- increase of the algorithmic
information content --- given by some $\Delta K(n) = K(\rho_S^{(n)}) -
K(\rho_S^{(o)})$.

The demon of choice would now either; (i) proceed with the expansion,
extraction and erasure, providing that his estimate of the future gain:
$$\Delta W = k_BT (\Delta H - \Delta K) = k_B T \Delta {\cal Z} \eqno (17)$$
was positive, or, alternatively; (ii) undo the measurement at no cost,
providing that $\Delta W < 0$. An algorithm that attempts to implement this
strategy for the case of Gabor's engine is illustrated in Fig. 3. To see why 
this strategy will not work, we first note that the demon of choice threatens
the second law only if its operation is cyclic --- that is, it must be
possible to implement the algorithm without it coming to an inevitable
halt.

There is no need to comment on the left-hand side part of the cycle: it
starts with the insertion of the partition. Detection of a particle
in the left-hand side compartment is
followed by the expansion of the partition (converted into a piston) and
results in extraction of $\Delta W^+_{p}$, Eq. (7), of work. Since
the partition was extracted, the results of the measurement must be erased
(to prepare for the next measurement) which costs $k_BT$ of useful work, so
that the gain per useful cycle is given by Eq. (9). The partition can be
now reinserted and the whole cycle can start again.

There is, however, no decision procedure which can implement the goal of
the right-hand side of the tree. The measurement can be of course undone.
The demon --- after undoing the correlation --- no longer knows 
the location of the molecule inside the engine. Unfortunately for the demon,
this does not imply that the state of the engine has also been undone. Moreover,
the demon with empty memory will immediately proceed to do what demons with
empty memory always do: It will measure. This action is an ``unconditional 
reflex" of a demon with an empty memory. It is inevitable, as the actions of 
the demon must be completely determined by its internal state, including
the state of its memory. (This is the same rule as for Turing machines.) 
But the particle in the Gabor's engine is still stuck on the unprofitable side 
of the partition. Therefore, when the measurement is repeated, it will yield 
the same disappointing result as before, and the demon will be locked forever 
into the measure - unmeasure ``two-step'' within the same unprofitable branch 
of the cycle by its algorithm, which compels it to repeat two controlled-not
like actions, Eq. (14), which jointly amount to an identity.

This vicious cycle could be interrupted only if the decision process called 
for extraction and reinsertion of the partition {\it before} undoing the 
measurement (and thus causing the inevitable immediate re-measurement) in the 
unprofitable right branch of the decision tree. Extraction of the partition 
before the measurement is undone increases the entropy of the gas by 
$k_B[\lg(L-\ell)/L]$ and destroys the correlation with the demon's memory, 
thus decreasing the mutual information: The molecule now occupies the whole 
volume of the engine. Moreover it occurs with no gain of useful work. 
Consequently, reversibly undoing the measurement {\it after} the partition is 
extracted is no longer possible: The location on the decision tree (extracted 
partition, ``full" memory) implicitly demonstrates that the measurement has 
been carried out and that it has revealed that the molecule was in the 
unprofitable compartment --- it can occurr only in the right hand branch of 
the tree.

The opening of the partition has resulted in a free expansion of the gas,
which squandered away the correlation between the state of the gas and the
state of the memory of the demon. Absence of the correlation eliminates the
possibility of undoing the measurement. Thus, now erasure is the only
remaining option. It would have to be carried out before the next
measurement, and the price of $k_B T$ per bit would have to be paid.$^{6,7}$

One additional strategy should be explored before we conclude this discussion: 
The demon of choice can be assumed to have a large memory tape, so that it can put 
off erasures and temporarily store the results of its ${\cal N}$ measurements.
The tape would then contain $\sim {\cal N} \cdot (\ell - L)/L$ 0's (which we 
shall take to signify an unprofitable outcome) and $\sim {\cal N} \ell/L$ 1's.
In the limit of large ${\cal N}$ $({\cal N}\ell/L>>1)$ the algorithmic 
information content of such a ``sparse'' binary sequence $s$ is given by$^{14-20}$:
$$K(s) \simeq - {\cal N} [{\ell \over L} \lg {\ell \over L} + {L - \ell
\over L} \lg {L - \ell \over L}] \eqno (18) $$
Moreover, a binary string can be, at least in principle, compressed to its
minimal record ($s^*$ such that $K(s) = \vert s^* \vert$) by a reversible 
computation.$^{12}$ Hence, it is possible to erase the record of
the measurements carried out by the demon at a cost of no less than
$$< \Delta W^- > = k_B T[K(s)/{\cal N}] \ . \eqno (19)$$
Thus, if the erasure is delayed so that the demon can attempt to
minimize its cost before carrying it out, it can at best break even: The
$-k_B T$ in Eq. (12) is substituted by the $< \Delta W^- >$, Eq. (19),
which yields:
$$< \Delta W > = < \Delta W^+ > + < \Delta W^- > = 0. \eqno (20)$$

It is straightforward to generalize this lesson derived on the example of
Gabor's engine to other situations. The essential ingredient is the
``noncommutativity'' of the two operations: ``undo the measurement" can be 
reversibly carried out only before ``extract the partition." The actions of 
the demon are, by the assumption of Church--Turing thesis, completely determined
by its internal state, especially its memory content. Demons are forced to make 
useless re-measurements. Famous Santayana's saying {\it those who forget their 
history are doomed to relive it} applies to demons with a vengance! 
For, when the demon forgets the measurement outcome, it will repeat the
measurement and remain stuck forever in the unprofitable cycle. One could
consider more complicated algorithms, with additional bits and instructions 
on when to measure, and so on. The point is, however, that all such strategies 
must ultimately contain explicit or implicit information about the branch
on which the demon has found itself as a result of the measurement. Erasure of 
this information carries a price which is on the average no less than the 
``illicit'' gains which would violate the second law.

The aim of this paper was to exorcise the demon of choice --- a selective
version of Maxwell's demon which attempted to capitalize on large
thermal fluctuations by reversibly undoing all of the measurements which
did not reveal the system to be sufficiently far from equilibrium. I have
demonstrated that a deterministic version of such a demon fails, as no
decision procedure is capable of both (i) reversibly undoing the measurement, 
and, also, of (ii) opening the partitions inserted prior to the measurement 
to allow for energy extraction following readoff of the outcome. 

Our discussion was phrased --- save for an occassional reference to density 
matrices, Hilbert spaces, etc. --- in a noncommital language, and it is indeed 
equally applicable in the classical and quantum contexts. 
As was pointed out already some time ago$^{9,10}$, the only difference
arises in the course of measurements. Quantum measurements are typically 
accompanied by a ``reduction of the state vector''. It ocurrs whenever observer
measures observables that are not co-diagonal with the density matrix of 
the system. It is a (near) instantaneous process$^{34}$, which is nowadays 
understood as a consequence of decoherence and einselection$^{28-34}$. 
The implications of this difference are minor from the viewpoint of the threat 
to the second law posed by the demons (although decoherence is paramount for the
discussion of the interpretation of quantum theory). It was noted already
some time ago that decoherence (or, more generally, the increase of entropy 
associated with the reduction of the state vector) is not necessary to save
the second law$^9$. Soon after the algorithmic information content entered
the discussion of demons$^{12,21}$ it was also realised that the additional cost
decoherence represents can be conveniently quantified using the ``deficit'' in 
what this author knew then as the `Gronewald--Lindblad inequality'$^{35,36}$, 
and what is now more often (and equally justifiably) called the 
`Holevo quantity;$^{37}$ 
$$ \chi = H(\rho) - \sum_i p_i H(\rho_S^{(i)}) \ , \eqno(21)$$
which is a measure of the entropy increase due to the ``reduction''. The two 
proofs$^{36,37}$ involving essentially the same quantity have appeared almost 
simultaneously, independently, and were motivated by --- at least superficially 
--- quite different considerations.

We shall not repeat these discussions here in detail. There are however several
independently sufficient reasons not to worry about decoherence in the demonic
context which deserve a brief review. To begin with, decoherence cannot help the
demon as it only adds to the ``cost of doing business''. And the second law is 
apparently safe even without decoherence$^9$. Moreover, especially in the 
context of Szilard's or Gabor's engines, decoherence is unlikely to hurt the 
demon either, since the obvious projection operators to use in Eq. (14) 
correspond to the particle being on the left (right) of the partition, and are 
likely to diagonalise the density matrix of the system in contact with a typical
environment$^9$ (heat bath). (Superpositions of states corresponding to such 
obvious measurement outcomes are very Schr\"odinger cat -- like, and, therefore,
unstable on the decoherence timescale$^{34}$.) Last not least, even if demon 
for some odd reason started by measuring some observable which does not commute
with the density matrix of the system decohering in contact with the heat bath
environment, it should be able to figure out what's wrong and learn after a 
while what to measure to minimise the cost of erasure (demons are supposed to
be intelligent, after all!).

So decoherence is of secondary importance in assuring validity of the second
law in the setting involving engines and demons: Entropy cannot decrease 
already without it! But decoherence can (and often will) add to the 
{\it measurement} costs, and the cost of decoherence is paid ``up front'',
during the measurement (and not really during the erasure, although there 
may be an ambiguity there --- see a quantum calculation of erasure -- like
process of the consequences of decoherence in Ref. 38). Moreover, in the context
of dynamics decoherence is the ultimate cause of entropy production, and,
thus, the cause of the algorithmic arrow of time$^{33}$. Moreover, there are 
intriguing quantum implication of the interplay of decoherence and (algorithmic)
information that follow: Discussions of the interpretational
issues of quantum theory are often conducted in a way which implicitly 
separates the information observers have about the state of the systems in the
``rest of the Universe'' from their own physical state --- their identity.
Yet, as the above analysis of the observer-like demons demonstrates, there
can be {\it no information without representation}. The observer's state (or, 
for that matter, the state of its memory) determines its actions and should be 
regarded as an ultimate description of its identity. So, to end with one more
``deep truth'' {\it existence} (of the observers state, and, especially, of 
the state of its memory) {\it precedes the essence} (observer's information,
and, hence their future actions).

I have benefited from discussion on this subject with many, including Andreas
Albrecht, Charles Bennett, Carlton Caves, Murray Gell-Mann, Chris
Jarzynski,
Demon Laflamme (who contributed to lowering entropy of the manuscript),
Rolf Landauer, Seth Lloyd, Michael Nielsen, Bill Unruh, and John Wheeler, who, 
in addition to stimulating the initial interest in matters concerning physics 
and information, insisted on my monthly dialogues with Feynman. This has led to 
one more ``adventure with a curious character'': In the Spring of 1984 
I participated in the ``Quantum Noise'' program at the Institute for Theoretical
Physics, UC Santa Barbara. It was to end with a one-week conference on various 
relevant quantum topics. One of the organisers (I think it was Tony Leggett), 
aware of my monthly escapades to Caltech, and of Feynman's (and mine) interests 
in quantum computation asked me whether I could ask him to speak. I did, and 
Feynman immediately agreed.

The lectures were held in a large conference room at the campus of the 
University of California at Santa Barbara. For the ``regular speakers'' and for 
most of the talks (such as my discussion of the decoherence timescale which 
was eventually published as Ref. 34) the room was filled to perhaps a third of 
the capacity. However, when I walked in in the middle of the afternoon coffee
break, well in advance of Feynman's
talk, the room was already nearly full, and the air was thick with anticipation.
A moment after I sat down in one of the few empty seats, I saw Feynman come in,
and quietly take a seat somewhere in the midst of the audience. More people
came in, including the organisers and the session chairman. The scheduled time
of his talk came... and went. It was five minutes after. Ten minutes. 
Quarter of an hour. The chairman was nervous. I did not understand what was
going on --- I clearly saw Feynman's long grey hair and an occasional flash of
an impish smile a few rows ahead. 

Then it struck me: He was just being ``a curious character'', curious about 
what will happen... He did what he had promised --- showed up for his talk on 
(or even before) time, and now he was going to see how the events unfold. 

In the end I did the responsible thing: After a few more minutes I pointed 
out the speaker to the session chairman (who was greatly relieved, and who
immediately and reverently led him to the speaker's podium). The talk 
(with the content, more or less, of Ref. 39) started only moderately behind 
the schedule. And I was immediately sorry that I did not play along a while 
longer --- I felt as if I had given away a high-school prank before it was 
fully consummated!

\vfill
\eject
\noindent{\bf References}

\item{1.} R. P. Feynman, R. B. Leighton, and M. Sands, {\it The Feynman Lectures
on Physics}, vol. 1, pp 46.1 -- 46.9 (Addison-Wesley, Reading, Massachussets, 
1963).

\item{2.} J. C. Maxwell, {\it Theory of Heat}, 4th ed., pp. 328-329
(Longman's, Green, \& Co., London 1985).

\item{3.} H. S. Leff and A. F. Rex, {\it Maxwell's Demon: Entropy, Information,
Computing}, (Princeton University Press, Princeton, 1990).

\item{4.} M. Smoluchowski in {\it Vortg\"age \"uber die Kinetische Theorie
der Materie und der Elektizit\"at} (Teubner, Leipzig 1914).

\item{5.} P. Skordos and W. H. Zurek, ``Maxwell's Demons, Rectifiers, and the
Second Law'' {\it Am. J. Phys.} {\bf 60}, 876 (1992).

\item{6.} L. Szilard, {\it Z. Phys.} {\bf 53} 840 (1929). English translation in
Behav. Sci. {\bf 9}, 301 (1964), reprinted in {\it Quantum Theory and
Measurement}, edited by J. A. Wheeler and W. H. Zurek (Princeton University
Press, Princeton, 1983); Reprinted in Ref. 3.

\item{7.} R. Landauer, {\it IBM J. Res. Dev.} {\bf 3}, 183 (1961); Reprinted 
in Ref. 3.

\item{8.} 
C. H. Bennett, {\it IBM J. Res. Dev.} {\bf 17} 525 (1973);
C. H. Bennett, {\it Int. J. Theor. Phys.} {\bf 21}, 905 (1982); 
C. H. Bennett, {\it IBM J. Res. Dev.}, {\bf 32}, 16-23 (1988); 
Reprinted in Ref. 3.

\item{9.} W. H. Zurek, ``Maxwell's Demons, Szilard's Engine's, and Quantum
Measurements", Los Alamos Preprint LAUR 84-2751 (1984); pp. 151-161 in 
{\it Frontiers of Nonequilibrium Statistical Physics}, G. T. Moore and 
M. O. Scully, eds., (Plenum Press, New York, 1986); reprinted in Ref. 3. 

\item {10.} For a quantum treatement which employs the Gronewold-Lindblad/Holevo
inequality and uses the ``deficit'' $\chi$ in that inequality to estimate of the
price of decoherence, see W. H. Zurek, pp 115-123 in the 
{\it Proceedings of the 3$^{rd}$ International 
Symposium on Foundations of Quantum Mechanics}, S. Kobayashi {\it et al.}, eds. 
(The Physical Society of Japan, Tokyo, 1990).

\item{11.} S. Lloyd, {\it Phys. Rev.} {\bf A56}, 3374-3382 (1997).

\item{12.} W. H. Zurek, {\it Phys. Rev.} {\bf A40}, 4731-4751 (1989);
W. H. Zurek, {\it Nature} {\bf 347}, 119-124 (1989).

\item{13.} For an accessible discussion of Church--Turing thesis, see
D. R. Hofstadter, {G\"odel, Escher, Bach}, chapter XVII (Vintage Books, 
New York, 1980).

\item{14.} R. J. Solomonoff, {\it Inf. Control} {\bf 7}, 1 (1964).

\item{15.} A. N. Kolmogorov, {\it Inf. Transmission} {\bf 1}, 3 (1965).

\item{16.} G. J. Chaitin, {\it J. Assoc. Comput. Mach.} {\bf 13}, 547 (1966).

\item{17.} A. N. Kolmogorov, {\it IEEE Trans. Inf. Theory} {\bf 14}, 662 (1968).

\item{18.} G. J. Chaitin, {\it J. Assoc. Comput. Mach.} {\bf 22}, 329 (1975);
G. J. Chaitin, {\it Sci. Am.} {\bf 23}(5), 47 (1975).

\item{20.} A. K. Zvonkin and L. A. Levin, {\it Usp. Mat. Nauk.} {\bf 25}, 602
(1970).

\item{21.} C. M. Caves, ``Entropy and Information'', pp. 91-116 in 
{\it Complexity, Entropy, and Physics of Information}, 
W. H. Zurek, ed. (Addison-Wesley, Redwood City, CA, 1990).

\item{22.} C. M. Caves, {\it Phys. Rev. Lett.} {\bf 64}, 2111-2114 (1990).

\item{23.} W. Shannon and W. Weaver, {\it  The Mathematical Theory of
Communication} (University of Illinois Press, Urbana, 1949).

\item{24.} C. M. Caves, W. G. Unruh, and W. H. Zurek, {\it Phys. Rev. Lett.},
to be supplied

\item{25.} D. Gabor, {\it Optics} {\bf 1}, 111-153 (1964).

\item{26.} L. Brillouin, {\it Science and Information Theory}, 2nd ed.
(Academic, London, 1962).

\item{27.} C. H. Bennett, {\it Sci. Am.} {\bf 255} (11), 108 (1987).

\item{28.} W. H. Zurek, {\it Phys. Rev.} {\bf D24}, 1516 (1981); {\it ibid.}
{\bf D26}, 1862 (1982); {\it Physics Today} {\bf 44}, 36 (1991).

\item{29.} W. H. Zurek, ``Information Transfer in Quantum Measurements: 
Irreversibility and Amplification'';  pp. 87-116 in {\it Quantum Optics, 
Experimental Gravitation, and Measurement Theory}, 
P. Meystre and M. O. Scully, eds. (Plenum, New York, 1983).

\item{30.} Joos, E. and Zeh, H. D., {\it Zeits. Phys.} {\bf B59}, 223 (1985).

\item{31.} Giulini, D., Joos, E., Kiefer, C., Kupsch, J., and Zeh, H. D., 
{\it Decoherence and the Appearance of a Classical World in Quantum Theory},
(Springer, Berlin, 1996). 

\item{32.} W. H. Zurek, {\it Progr. Theor. Phys.} {\bf 89}, 281-312 (1993).
 
\item{33.} W. H. Zurek, in the {\it Proceedings of the Nobel Symposium 101 
`Modern Studies in Basis Quantum Concepts and Phenomena'}, to appear in 
{\it Physica Scripta}, in press {\tt quant-ph/9802054}.

\item{34.} W. H. Zurek, ``Reduction of the Wavepacket: How Long Does it Take?"
Los Alamos preprint LAUR 84-2750 (1984);
pp. 145-149 in the {\it Frontiers of Nonequilibrium Statistical Physics: 
Proceedings of a NATO ASI held June 3-16 in Santa Fe, New Mexico},
G. T. Moore and M. O. Scully, eds. (Plenum, New York, 1986).

\item{35.} H. J. Groenwold, {\it Int. J. Theor. Phys.} {\bf 4}, 327 (1971).

\item{36.} G. Lindblad, {\it Comm. Math. Phys.} {\bf 28}, 245 (1972).

\item{37.} A. S. Holevo, {\it Problemy Peredachi Informatsii} {\bf 9}, 9-11 
(1973).

\item{38.} J. R. Anglin, R. Laflamme, W. H. Zurek, and J. P. Paz, 
{\it Phys. Rev.} {\bf D52}, 2221-2231 (1995).

\item{39.} R. P. Feynman, ``Quantum Mechanical Computers'', {\it Optics News},
reprinted in {\it Found. Phys.} {\bf 16}, 507-531 (1986).

\vfill
\eject

\noindent Figure Captions:

\bigskip
\noindent Fig. 1 Gabor's engine.$^{25}$. See text for the standard operating
procedure. The decision between the two branches (of which only one --- the 
profitable one --- is shown) can be made reversibly with the help of the device
shown in Fig. 2.
\medskip
\noindent Fig. 2 Blueprint of a reversible measuring device for Gabor's engine.
The measurements can be done (or undone) by turning the crank on the right in 
the appropriate direction and pushing in or pulling out the ``scale''. 
Thermodynamic 
reversibility is achieved in the limit of an infinitesimally slow operation. 
Faster controlled-not like measurements can be carried out on a dynamical 
timescale by implementing the unitary evolution given by Eq. (14). The design 
shown above is similar to the Szilard's engine contraption devised in Ref. 28.
\medskip
\noindent Fig. 3 Decision flowchart for the demon of choice. The branch on the
left is profitable (and it is followed when the particle is ``caught'' in the
small left chamber, see Fig. 1). The branch on the right is unprofitable, and
as it is explained in the text in more detail, the demon of choice cannot be 
``saved'' by reversing only the unprofitable measurements. 

 \epsfxsize= 5truein \epsfbox{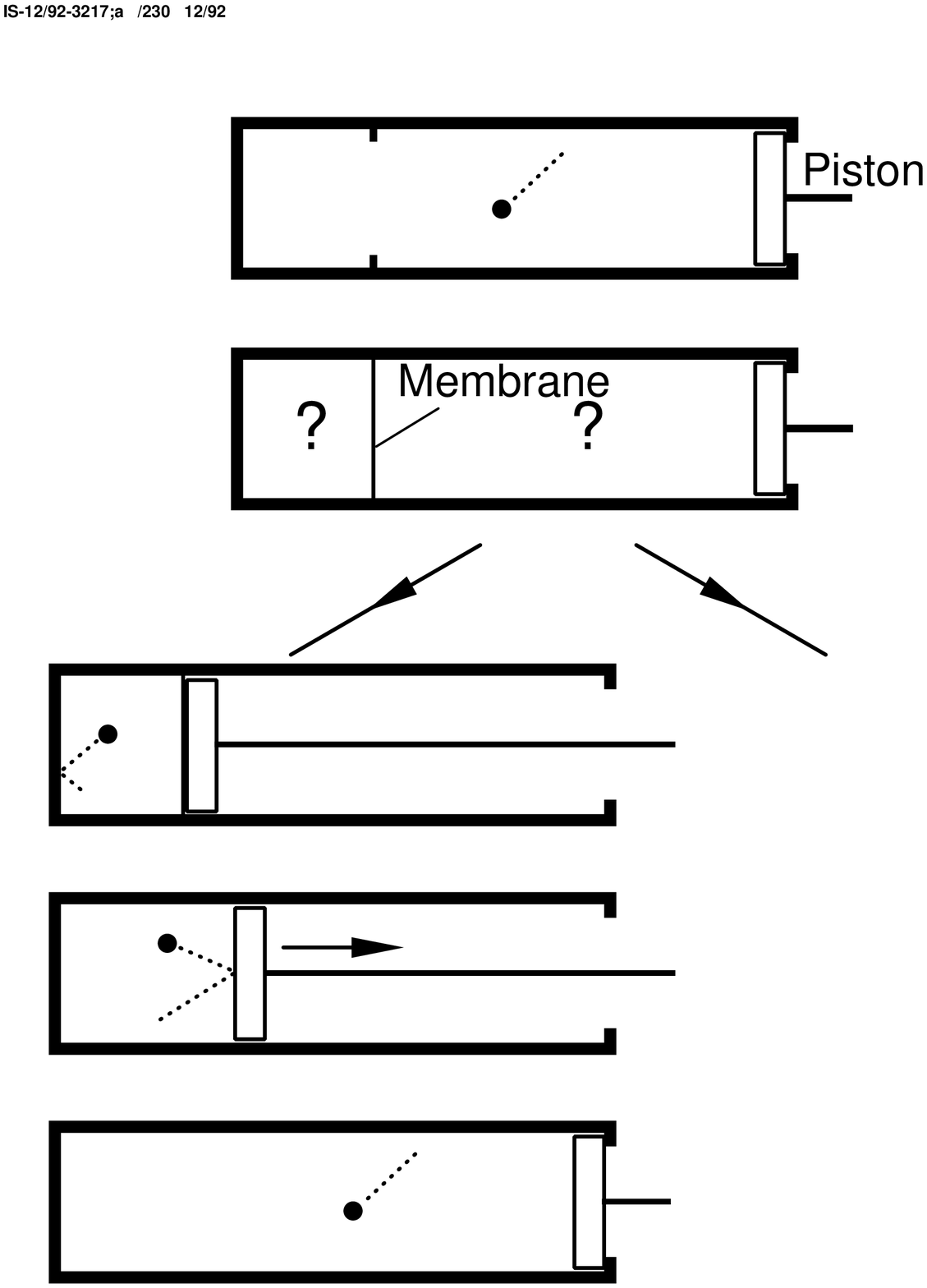} 

 \epsfxsize= 5truein\epsfbox{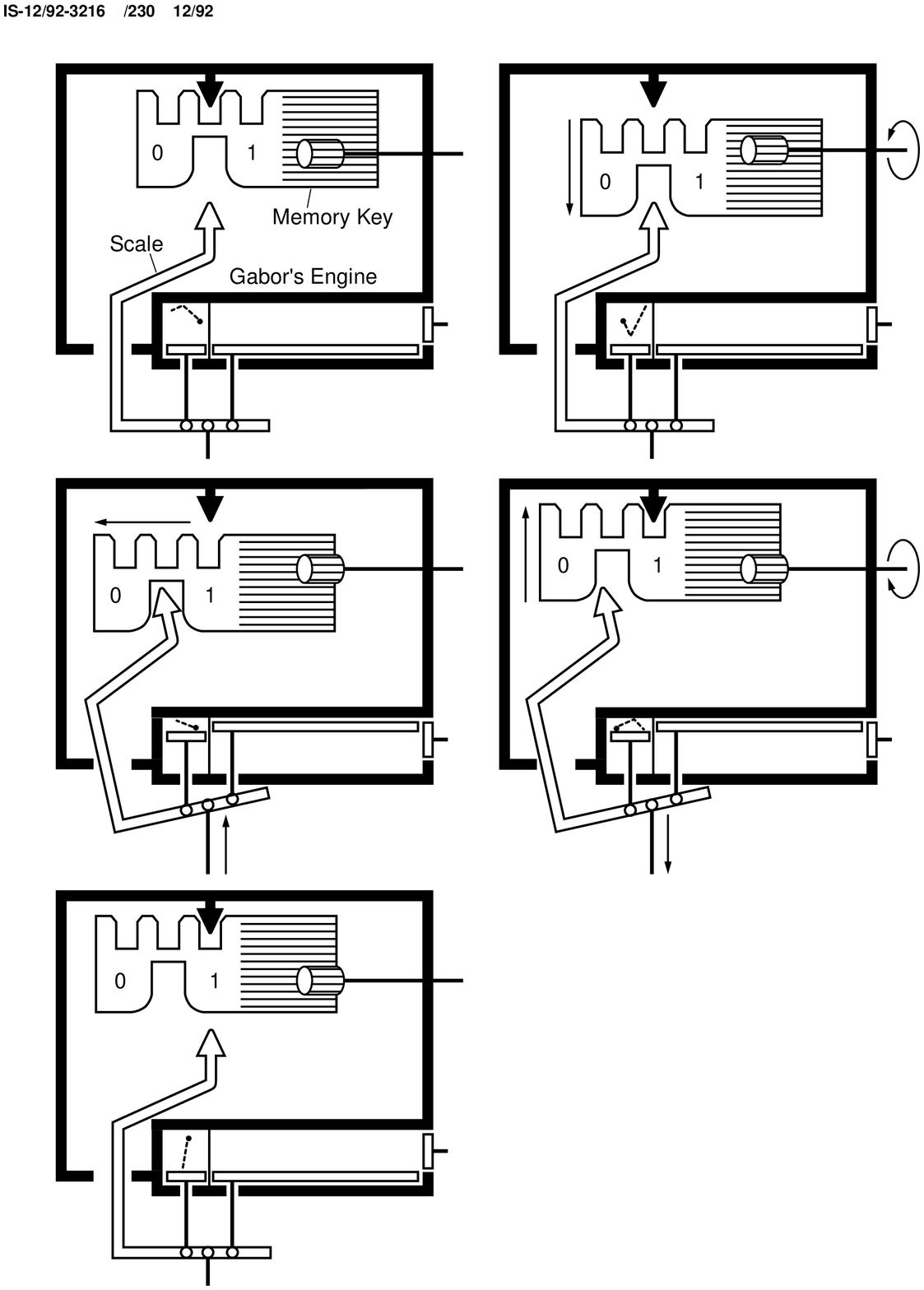} 

 \epsfxsize= 5truein \epsfbox{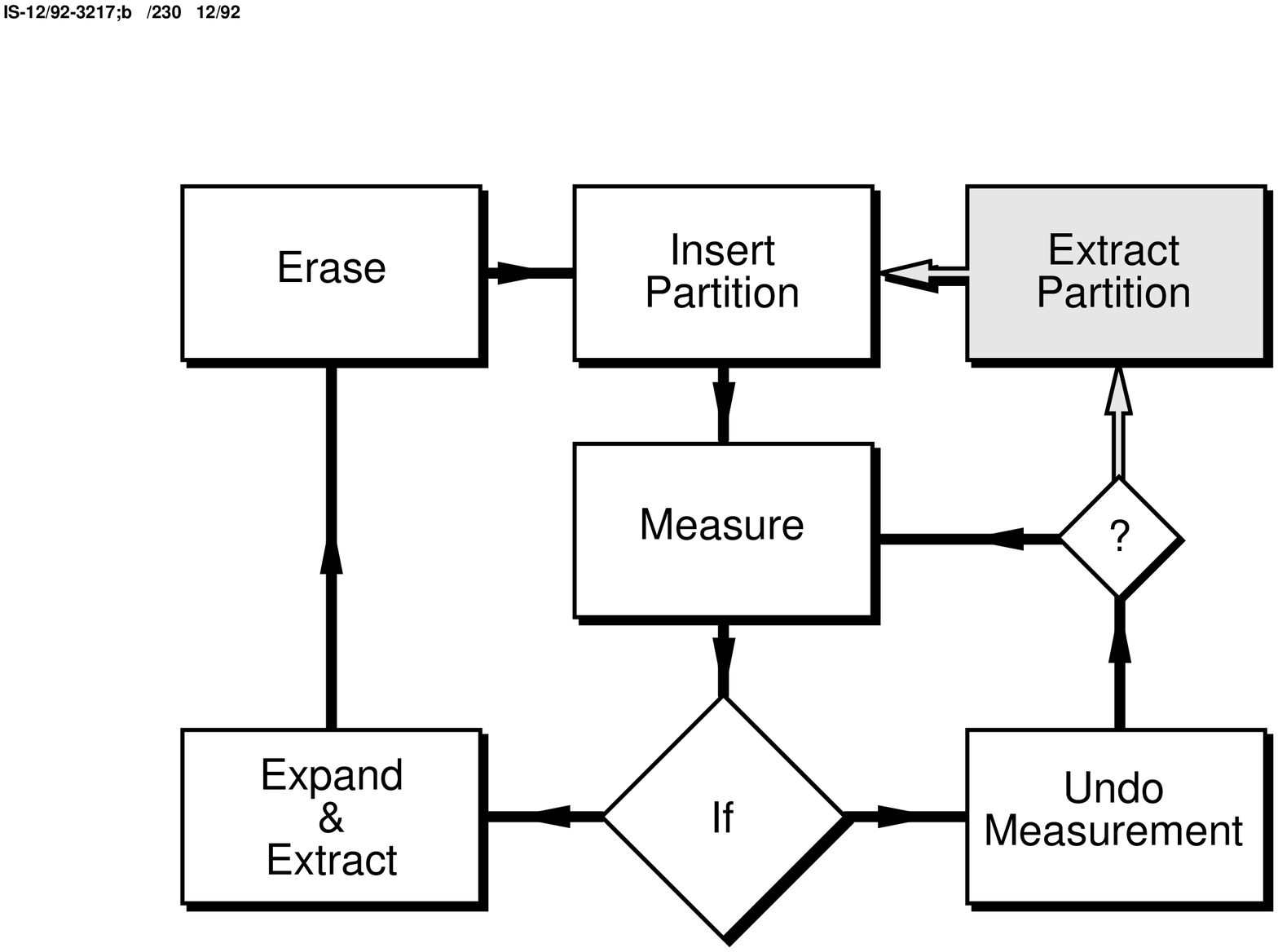} 

\end